\documentclass[preprint,prb]{revtex4}
\usepackage{graphicx}
\usepackage{latexsym}
\usepackage{amsfonts}
\usepackage{sparticles}
\usepackage{feynmf}
\begin{document}

\title{Effective Quantum Dynamics of Quarks and "Gluons" in a Stochastic Background}
\author{Jose A. Magpantay}
\email{jose.magpantay@up.edu.ph}

\affiliation{ National Institute of Physics, University of the
Philippines, Diliman Quezon City, 1101,Philippines }

\date{\today}

\begin{abstract}
The quantum dynamics of quarks and "gluons" and the scalar degrees
of freedom associated with the non-linear regime of the non-linear
gauge is derived. We discuss the subtleties in quantizing in a
stochastic background. Then we show in detail that $\langle
S_{YM}(t^{a}_{\mu},\phi^{a};\tilde{f}^{a})\rangle_{\tilde{f}}$
only depends on $t^{a}_{\mu}$, thus effectively proving that the
scalars $\phi^{a}$ are non-propagating.  Integrating out the
scalars from $\langle S_{fermion}\rangle_{\tilde{f}}$ leads to
fermion wave function that declines exponentially. Finally, we
derive the effective action of the "gluons" and fermions resulting
from stochastic averaging. We show that it leads to a confining
four-fermi interaction.
\end{abstract}

\maketitle

\section{Introduction}

The non-linear gauge condition\cite{1}
\begin{equation}\label{eq:1}
(\partial\cdot D^{ab})(\partial\cdot
A^{b})=(D^{ab}\cdot\partial)(\partial\cdot
A^{b})=(\partial^{2}\delta^{ab}-g\epsilon^{abc}A^{c}_{\mu}\partial_{\mu})(\partial\cdot
A^{b})=0,
\end{equation}
and its physical consequences has been discussed by this author in
a number of papers.  This gauge condition is a natural
generalization in Yang-Mills theory of the Coulomb gauge in
Abelian theory. The reasons for this claim are:

(a) In the Abelian limit, $(\partial\cdot
D)\rightarrow\partial^{2}$ and since $\partial^{2}$ is positive
definite, the gauge-condition yields the Coulomb gauge.

(b) The gauge condition has two regimes, the linear regime given
by $\partial\cdot A^{a}=0$, the Coulomb gauge, and the non-linear
regime defined by $\partial\cdot A^{a}=f^{a}\neq 0$, which is also
the zero mode of $\partial\cdot D$.  Thus the non-linear regime
corresponds to the "Gribov" horizon on the surface $\partial\cdot
A^{a}=f^{a}(x)$.

(c) The linear and non-linear regimes do not mix in the sense that
field configurations in the non-linear regime cannot be gauge
transformed to the Coulomb gauge and vice versa.\cite{2}

(d) If we consider the running of the coupling constant, we find
that for the short-distance regime where $g\rightarrow 0$
(asymptotic freedom phase), the non-linear gauge reduces to
$\partial^{2}(\partial\cdot A^{a})=0$, yielding the Coulomb gauge
because $\partial^{2}$ is positive definite. This is consistent
with the fact that transverse gluons are physical degrees of
freedom at short-distances. However, as we increase the distance
scale, $g$ increases and before the running coupling becomes too
large, where perturbation theory loses validity, the full
non-linear character of the gauge condition becomes important
where the relevant degrees of freedom are the scalars $f^{a}(x)$
and the "gluons" $t^{a}_{\mu}(x)$, which decompose the Yang-Mills
potential\cite{3} as
\begin{equation}\label{eq:2}
A^{a}_{\mu}=\frac{1}{(1+\vec{f}\cdot\vec{f})
}(\delta^{ab}+\epsilon^{abc}f^{c}+f^{a}f^{b})(\frac{1}{g}\partial_{\mu}f^{b}+t^{b}_{\mu}).
\end{equation}
This means that the non-linear gauge continuously interpolates
short-distance and large distance physics with their corresponding
degrees of freedom.  Since we consider quantum effects in the
running of the coupling, the non-linear gauge is a quantum gauge
condition.

(e) Lastly, geometrically, configuration space analysis showed
that non-linear gauge condition is a modification of the global
orthogonal gauge condition, which does not exist in non-Abelian
theories and gives the Coulomb gauge in Abelian theory.\cite{4}

As for the physical consequences of the non-linear regime of the
non-linear gauge, the following results had been established by
the author:

(a) The Yang-Mills action is quartic in $t^{a}_{\mu}$ and
infinitely non-linear in $f^{a}$.  The pure $f^{a}$ action has a
$\frac{1}{g^{2}}$ factor and its kinetic term goes like $(\partial
f^{a})^{4}$.  Clearly, these hint of non-perturbative
physics.\cite{5}

(b) The pure $f^{a}$ dynamic shows that all spherically symmetric
$\tilde{f}^{a}(x)$ with $x=(x_{\mu}x_{\mu})^{1/2}$ are classical
configurations with zero field strength.  The pure $f^{a}$
dynamics has a very broad minimum with zero action.  Because of
the infinite degeneracy in classical configurations, the author
proposed to treat $\tilde{f}^{a}$ as a stochastic variable with a
white-noise distribution.  This resulted in the area law behaviour
of the Wilson loop, which means a linear potential for static
sources.\cite{3}

(c) If the stochastic treatment of the classical configuration
$\tilde{f}^{a}(x)$ yields a linear potential between static
sources, full quantum dynamics of $f^{a}$ shows equivalence to an
$O(1,3)$ non-linear $\sigma$ model in 2D.\cite{6} The proof made
use of the Parisi-Sourlas mechanism.\cite{7} Since the $\sigma$
model is $O(1,3)$ and not $O(4)$ resulting in a kinetic term with
a wrong sign, the proof of confinement is purely formal.
Furthermore, the proof of confinement only involves the scalars
and not the "gluons" and quarks.

(d) When we consider the classical dynamics of the "gluons" in the
spherically symmetric background $\tilde{f}^{a}(x)$, it was shown
that stochastically averaging $\tilde{f}^{a}(x)$ yielded a mass
for the $t^{a}_{\mu}$ and the loss of its
self-interactions.\cite{8} Thus, we have shown the mechanism for
the mass gap.  As for the loss of self-interactions, the result
was also arrived at by Kondo.\cite{9}

(e) Finally, the author proposed the concept of limited gauge
invariance and showed that in both the linear and non-linear
regimes of the non-linear gauge we can define potentials that are
gauge-invariant within their limited context.\cite{8}

In this paper, we will address the shortcomings noted in (c) by
considering quarks and "gluons" in a stochastic background.  We
will decompose the scalar $f^{a}(x)$ via
\begin{equation}\label{eq:3}
f^{a}(x)=\tilde{f}^{a}(x)+\phi^{a}(x).
\end{equation}
We will note the subtleties in treating quantum and stochastic
fluctuations. We will derive the effective action of quarks and
"gluons" by averaging over the stochastic background. Finally, we
will derive a confining non-local four-fermi interaction.

\section{The Action in the Non-Linear Gauge}

We will consider $SU(2)$ theory with the following action
\begin{equation}\label{eq:4}
S=S_{YM}+S_{fermion}=\int
d^{4}x\{\frac{1}{4}F^{a}_{\mu\nu}F^{a}_{\mu\nu}+\bar{\psi}i\gamma_{\mu}D_{\mu}\psi\}.
\end{equation}
Introducing the Fadeev-Popov trick of resolving unity using the
gauge condition given by equation (\ref{eq:1}), we find the vacuum
to vacuum functional
\begin{equation}\label{eq:5}
W(0)=\int
(dA^{a}_{\mu})(d\psi)(d\bar{\psi})det\theta\delta(\partial\cdot
D(\partial\cdot A))e^{-S},
\end{equation}
where
\begin{equation}\label{eq:6}
\theta^{ad}=(D\cdot^{ab}\partial)(\partial\cdot
D^{bd})-g\epsilon^{abc}(\partial_{\mu}(\partial\cdot
A^{b}))D^{cd}_{\mu}.
\end{equation}
Note that in the linear regime, where the coupling $g$ is very
weak and thus the smallest eigenvalue of the positive definite
$\partial^{2}$ (in $R^{4}$) is not lowered to zero,
$(\partial\cdot D)$ is still positive definite and the
path-integral  given by equation (\ref{eq:5}) reduces to the
transverse gauge path-integral.

As we further increase  the coupling constant, we get to the
Gribov horizon of $\partial\cdot A^{a}=0$ surface.  We are still
on the Coulomb surface but the Fadeev-Popov operator is now
singular. This regime had been discussed extensively in the late
1970's and 1980s.  The conjecture is that restricting the
transverse gauge path-integral to within the central Gribov
region\cite{10} or the fundamental modular region results in
confinement\cite{11}. This author has a different view:
confinement happens in the next stage.

Further increasing the distance scale, we get to the regime where
transverse gluons should no longer be relevant degrees.  The
reasons for this are the existence of the mass gap (massive vector
fields are not transverse) and the fact that it is essentially
impossible to get a confining interaction from the exchange of
transverse gluons. Thus, we should be getting off the Coulomb
surface and the non-linear gauge seems to be a natural gauge to
consider.  Here, we get to the Gribov horizon of the
$\partial\cdot A^{a}=f^{a}\neq 0$ surface where the zero mode of
$(\partial\cdot D)$ is $\partial\cdot A^{a}$ itself. However, even
though $\partial\cdot D$ is singular, the corresponding operator
$\theta$ given by equation (\ref{eq:6}) is non-singular\cite{4}.
The path-integral is still well-defined, presenting no need to
deal with zero modes.

In this non-linear regime, we can decompose $A^{a}_{\mu}$ in terms
of the scalars $f^{a}$ and "gluons" $t^{a}_{\mu}$ as given in
equation (\ref{eq:2}).  These new degrees of freedom satisfy the
constraints
\begin{eqnarray}
\partial\cdot t^{a}-\frac{1}{g\ell^{2}}f^{a}=0\label{eq:7}\\
\rho^{a}=\frac{1}{(1+\vec{f}\cdot\vec{f})^{2}}[\epsilon^{abc}+\epsilon^{abd}f^{d}f^{c}-\epsilon^{acd}f^{d}f^{b}+f^{a}f^{d}\epsilon^{dbc}\nonumber\\
-f^{a}(1+\vec{f}\cdot\vec{f})\delta^{bc}-f^{c}(1+\vec{f}\cdot\vec{f})\delta^{ab}]\partial_{\mu}f^{b}t^{c}_{\mu}=0,\label{eq:8}
\end{eqnarray}
giving the same number of degrees of freedom as the original
$A^{a}_{\mu}(x)$.  Substituting equation (\ref{eq:2}) in equation
(\ref{eq:4}), we get
\begin{eqnarray}\label{eq:9}
S=\int d^{4}x\{\langle\frac{1}{g^{2}}Z^{2}(f)+\frac{2}{g}Z(f)\cdot
L(f,t)+[2Z(f)\cdot Q(f,t)+L(f,t)\cdot L(f,t)]\nonumber\\
+2gL(f,t)\cdot Q(f,t)+g^{2}Q(f,t)\cdot Q(f,t)\rangle
+\langle\bar{\psi}i\gamma_{\mu}\partial_{\mu}\psi\nonumber\\
-ig \bar{\psi}\gamma_{\mu}T^{a}\psi
[R^{ab}(f)(\frac{1}{g}\partial_{\mu}f^{b}+t^{b}_{\mu})]\rangle\},
\end{eqnarray}
where
\begin{eqnarray}
Z^{a}_{\mu\nu}(f)=X^{abc}(f)\partial_{\mu}f^{b}\partial_{\nu}f^{c},\label{eq:10}\\
L^{a}_{\mu\nu}(f,t)=R^{ab}(f)(\partial_{\mu}t^{b}_{\nu}-\partial_{\nu}t^{b}_{\mu})+Y^{abc}(\partial_{\mu}
f^{b}t^{c}_{\nu}-\partial_{\nu}f^{b}t^{c}_{\mu}),\label{eq:11}\\
Q^{a}_{\mu\nu}(f,t)=T^{abc}(f)t^{b}_{\mu}t^{c}_{\nu},\label{eq:12}\\
X^{abc}(f)=\frac{1}{(1+\vec{f}\cdot\vec{f})^{2}}[-(1+2\vec{f}\cdot\vec{f})\epsilon^{abc}+2\delta^{ab}f^{c}-2\delta^{ac}f^{b}\nonumber\\
+
3\epsilon^{abd}f^{d}f^{c}-3\epsilon^{acd}f^{d}f^{b}+\epsilon^{bcd}f^{a}f^{d}],\label{eq:13}\\
R^{ab}(f)=\frac{1}{(1+\vec{f}\cdot\vec{f})}(\delta^{ab}+\epsilon^{abc}f^{c}+f^{a}f^{b}),\label{eq:14}\\
Y^{abc}=\frac{1}{(1+\vec{f}\cdot\vec{f})^{2}}[-(\vec{f}\cdot\vec{f})\epsilon^{abc}+(1+\vec{f}\cdot\vec{f})f^{a}\delta^{bc}-(1-\vec{f}\cdot\vec{f})\delta^{ac}f^{b}\nonumber\\
+
3\epsilon^{cad}f^{d}f^{b}-2f^{a}f^{b}f^{c}+\epsilon^{abd}f^{d}f^{c}+f^{a}\epsilon^{bcd}f^{d}],\label{eq:15}\\
T^{abc}=\frac{1}{(1+\vec{f}\cdot\vec{f})^{2}}[\epsilon^{abc}+(1+\vec{f}\cdot\vec{f})f^{b}\delta^{ac}-(1+\vec{f}\cdot\vec{f})f^{c}\delta^{ab}\nonumber\\
+\epsilon^{abd}f^{d}f^{c}+f^{a}\epsilon^{bcd}f^{d}+\epsilon^{acd}f^{d}f^{b}]\label{eq:16}.
\end{eqnarray}

The pure $f^{a}$ dynamics given by $Z^{2}$ has a class of
classical configurations with zero field strength: the spherically
symmetric $\tilde{f}^{a}(x)$, with $x=(x_{\mu}x_{\nu})^{1/2}$ (we
are in $R^{4}$). This follows from the anti-symmetry of $X^{abc}$
with respect to the last two indices and that for spherically
symmetric $\tilde{f}^{a}(x)$,
$\partial_{\mu}\tilde{f}^{a}(x)=\frac{x_{\mu}}{x}\frac{d\tilde{f}^{a}}{dx}$.
Since we will be expanding about a class of classical
configurations, the author proposed to treat the spherically
symmetric $\tilde{f}^{a}(x)$ stochastically with a white-noise
distribution.

What will be done in the remainder of this section is to
substitute the background decomposition given by equation
(\ref{eq:3}) in the action given by equations (\ref{eq:9}) and
(\ref{eq:10}) to (\ref{eq:16}). The resulting action is quantic in
$t^{a}_{\mu}$ and infinitely non-linear in $\phi^{a}$ with
coefficients that are functions of $\tilde{f}^{a}(x)$. The result
is (see Appendix A).
\begin{eqnarray}\label{eq:17}
S&=&\int
d^{4}x\{[K^{ab}_{\mu\nu}(\tilde{f})\partial_{\mu}\phi^{a}\partial_{\nu}\phi^{b}
+M^{ab}(\tilde{f})\phi^{a}\phi^{b}+N^{ab}_{\mu}(\tilde{f})\partial_{\mu}\phi^{a}\phi^{b}\nonumber\\
&+&\sum^{\infty}_{n=3}\langle I^{a_{1}\cdots
a_{n}}(\tilde{f})\phi^{a_{1}}\phi^{a_{2}}+J^{a_{1}\cdots
a_{n}}(\tilde
{f})\phi^{a_{1}}\partial_{\mu}\phi^{a_{2}}+H^{a_{1}\cdots
a_{n}}\partial_{\mu}\phi^{a_{1}}\partial_{\nu}\phi^{a_{2}}\rangle\nonumber\\
&\times & \phi^{a_{3}}\cdots
\phi^{a_{n}}]+\frac{1}{4}[\mathbb{R}^{ab}(\tilde{f})(\partial_{\mu}t^{a}_{\nu}-\partial_{\nu}t^{a}_{\mu})
(\partial_{\mu}t^{b}_{\nu}-\partial_{\nu}t^{b}_{\mu})\nonumber\\
&+&2\mathbb{S}^{ab}(\tilde{f})(\partial_{\mu}t^{a}_{\nu}-\partial_{\nu}t^{a}_{\mu})
(\frac{x_{\mu}}{x}t^{b}_{\nu}-\frac{x_{\nu}}{x}t^{b}_{\mu})
+2\mathbb{Y}^{ab}(\tilde{f})t^{a}_{\mu}t^{b}_{\nu}\nonumber\\
&+& 2g\mathbb{U}^{abc}(\partial_{\mu}t^{a}_{\nu}-\partial_{\nu}t^{a}_{\mu})t^{b}_{\mu}t^{c}_{\nu}
+g^{2}\mathbb{T}^{abcd}t^{a}_{\mu}t^{b}_{\mu}t^{c}_{\nu}t^{d}_{\nu}]\nonumber\\
&+&\frac{1}{4}\sum^{\infty}_{n=1}\frac{1}{n!}\frac{\delta^{n}}{\delta
f^{a_{1}}(x_{2})\cdots\delta
f^{a_{n}}(x_{n})}[\frac{2}{g}Z(f)\cdot
L(f,t)+L^{2}(f,t)+2Z(f)\cdot Q(f,t)\nonumber\\
&+& gL(f,t)\cdot
Q(f,t)+g^{2}Q(f,t)Q(f,t)]_{f=\tilde{f}}\phi^{a_{1}}(x_{1})\cdots\phi^{a_{n}}(x_{n})\nonumber\\
&+&[\bar{\psi}i\gamma_{\mu}\partial_{\mu}\psi +
\bar{\psi}i\gamma_{\mu}T^{a}\langle
R^{ab}(\tilde{f})(\frac{1}{g}\partial_{\mu}\tilde{f}^{b}+\frac{1}{g}\partial_{\mu}\phi^{b}+t^{b}_{\mu})\nonumber\\
&+&
\sum^{\infty}_{n=1}\frac{1}{n!}\frac{\delta^{n}R^{ab}(\tilde{f})}{\delta
f^{c_{1}}(x_{1})\cdots\delta
f^{c_{n}}(x_{n})}(\frac{1}{g}\partial_{\mu}\tilde{f}^{b}+\frac{1}{g}\partial_{\mu}
\phi^{b}+t^{b}_{\mu})\phi^{c_{1}}(x_{1})\cdots\phi^{c_{n}}(x_{n})\rangle
\psi (x)]\}.
\end{eqnarray}

The first term bracketted with [ ] comes from $Z^{2}(f)$ where we
made use of $Z^{a}_{\mu\nu}(\tilde{f})=0$. The second bracketted
terms have been discussed in a previous paper \cite{9}, where the
author presented the classical dynamics of "gluons" in a
stochastic vacuum.  The third shows the interactions between the
"gluons" and scalars $\phi^{a}$. The fourth gives the fermion
action, with the interaction between the fermions and "gluons" and
scalars in a classical background $\tilde{f}^{a}(x)$. Equation
(\ref{eq:17}) looks very complicated where the scalars are
infinitely non-linear and the background $\tilde{f}^{a}(x)$ is
rather involved. However, this complicated action is dramatically
simplified upon stochastic averaging.

Introduce the white noise distribution
\begin{equation}\label{eq:18}
P[\tilde{f}]=\mathcal{N}
exp.\{-\frac{1}{\ell}\int^{\infty}_{0}\tilde{f}^{a}(s)\tilde{f}^{a}(s)ds\}
\end{equation}
where $\ell$ is the scale when non-perturbative physics is
important. From the running of the coupling, we must have
$\ell\sim\Lambda^{-1}_{QCD}$. When we average equation
(\ref{eq:17}) using equation (\ref{eq:18}), a tremendous
simplification results (see Appendix B)
\begin{eqnarray}\label{eq:19}
\langle
S\rangle_{\tilde{f}}&=&\int(d\tilde{f}^{a}(x))S(t^{a}_{\mu},\phi^{a},\psi,\bar{\psi};\tilde{f})P[\tilde{f}]\nonumber\\
&=&\frac{1}{4}\int
d^{4}x\{\frac{1}{3}(\partial_{\mu}t^{a}_{\nu}-\partial_{\nu}t^{a}_{\mu})^{2}+\frac{3}{2}(\frac{n}{\ell})^{2}
t^{a}_{\mu}t^{a}_{\mu}+\bar{\psi}i\gamma_{\mu}[\partial_{\mu}-gT^{a}(\frac{1}{3}t^{a}_{\mu}\nonumber\\
&+&\frac{1}{3g}\partial_{\mu}\phi^{a}-\frac{2}{3g}(\frac{n}{\ell})\frac{x_{\mu}}{x}\phi^{a})]\psi\}.
\end{eqnarray}
From equation (\ref{eq:19}), we see that the scalars $\phi^{a}$ do
not have a kinetic term; they do not propagate.

When we get the equation of motion for $\phi^{a}$, we find
\begin{equation}\label{eq:20}
\partial_{\mu}\psi +\frac{1}{(\frac{\ell}{n})}g^{2}\frac{x_{\mu}}{x}\psi = 0.
\end{equation}
The spherically symmetric solution of equation (\ref{eq:20}) is
\begin{equation}\label{eq:21}
\psi(x)=\psi_{0} exp.\{-\frac{1}{(\frac{\ell}{n})}g^{2}x\},
\end{equation}
which is an exponentially vanishing fermion field with effective
length $=\frac{\ell}{ng^{2}}\sim\Lambda^{-1}_{QCD}$. This
behaviour is a signal of confinement.

\section{Quantum Field Theory in a Stochastic Background}

We would like to quantize $t^{a}_{\mu},\psi,\bar{\psi}$ and
$\phi^{a}$ in the presence of the stochastic background
$\tilde{f}^{a}(x)$. In essence, we need to deal with quantum
fluctuations on top of classical stochastic variables. This poses
an ambiguity due to the ordering of how these two fluctuations are
taken into account. One sequence is to consider quantum field
theory in a stochastic background and then do the stochastic
averaging.  The other sequence is first to do the stochastic
averaging on the classical action and then quantize the resulting
theory.

Let us collectively represent the fields
$t^{a}_{\mu},\psi,\bar{\psi}$ and $\phi^{a}$ by the field $\Phi$.
The stochastic classical background is still given as $\tilde{f}$.
We write the action of $\Phi$ in a background $\tilde{f}$, which
includes gauge-fixing and the "Fadeev-Popov" determinant as
$S[\Phi,\tilde{f}]=\int d^{4}x\mathcal{L}(\Phi,\tilde{f})$. Define
\begin{eqnarray}\label{eq:22}
\langle
S[\Phi,\tilde{f}\rangle_{\tilde{f}}&\equiv&\int(d\tilde{f})\mathcal{N}
e^{-\frac{1}{\ell}\int^{\infty}_{0}\tilde{f}(s)\tilde{f}(s)}S[\Phi,\tilde{f}]\nonumber\\
&=& S_{eff}[\Phi].
\end{eqnarray}
The complete generating functional from which we compute n-point
functions is
\begin{equation}\label{eq:23}
W_{eff}[J]=\int(d\Phi)e^{-S_{eff}[\Phi]-\int d^{4}x J\Phi}
\end{equation}
From this generating functional, we compute the n-point function
by
\begin{equation}\label{eq:24}
G_{n}(x_{1},\cdots, x_{n})=\frac{\delta^{n}W_{eff}[J]}{\delta
J(x_{1})\cdots\delta J(x_{n})}.
\end{equation}

On the other hand, we can first consider the complete generating
functional in the presence of the background $\tilde{f}$. This is
\begin{equation}\label{eq:25}
W[J;\tilde{f}]=\int(d\Phi)e^{-S[\Phi,\tilde{f}]-\int d^{4}x
J\Phi}.
\end{equation}
Doing stochastic averaging, we find
\begin{eqnarray}\label{eq:26}
W'_{eff}[J]&=&\langle W[J,\tilde{f}]\rangle_{\tilde{f}}\nonumber\\
&=&\int(d\Phi)\langle
e^{-S[\Phi,\tilde{f}]}\rangle_{\tilde{f}}e^{-\int d^{4}xJ\Phi}
\end{eqnarray}

Expanding the exponential and taking the stochastic average of
each term, we find
\begin{equation}\label{eq:27}
\langle
e^{-S[\Phi,\tilde{f}]}\rangle_{\tilde{f}}=e^{-S'_{eff}[\Phi]},
\end{equation}

where
\begin{eqnarray}\label{eq:28}
S'_{eff}[\Phi]&=&\langle
S[\Phi,\tilde{f}]\rangle_{\tilde{f}}-\frac{1}{2}\int d^{4}x d^{4}y
\langle\underbrace{\mathcal{L}(\Phi,\tilde{f};x)\mathcal{L}(\Phi,\tilde{f};y)}\rangle\nonumber\\
&-&\frac{1}{4!}\int d^{4}x d^{4}y d^{4}z d^{4}r
\langle\underbrace{\mathcal{L}(\Phi,\tilde{f};x)\mathcal{L}(\Phi,\tilde{f};y)\mathcal{L}(\Phi,\tilde{f};z)\mathcal{L}(\Phi,\tilde{f};r)}\rangle_{\tilde{f}}\nonumber\\
&+&\cdots
\end{eqnarray}

The terms
$\underbrace{\mathcal{L}(\Phi,\tilde{f};x)\mathcal{L}(\Phi,\tilde{f};y)}$,
etc., are correlated points, which arise because the derivative of
the white noise $\tilde{f}(x)$ is "smoothened-out" via
\begin{equation}\label{eq:29}
\frac{d\tilde{f}^{a}}{dx}=\frac{\tilde{f}^{a}(x+\frac{\ell}{n})-\tilde{f}^{a}(x)}{\frac{\ell}{n}}
\end{equation}
Using equations (\ref{eq:27}) and (\ref{eq:28}) in (\ref{eq:26}),
we find that $W'_{eff}[J]\neq W_{eff}[J]$. The n-point function
before the stochastic averaging is given by
\begin{equation}\label{eq:30}
G_{n}(x_{1}\cdots,
x_{n};\tilde{f})=\frac{\delta^{n}W[J;\tilde{f}]}{\delta
J(x_{1})\cdots\delta J(x_{n})}.
\end{equation}
Obviously,
\begin{eqnarray}\label{eq:31}
\langle G_{n}(x_{1},\cdots,
x_{n};\tilde{f})\rangle_{\tilde{f}}&=&\langle\frac{\delta^{n}W[J;\tilde{f}]}{\delta
J(x_{1})\cdots\delta J(x_{n})}\rangle_{\tilde{f}}\nonumber\\
&=& \frac{\delta^{n}W'_{eff}[J]}{\delta J(x_{1})\cdots\delta
J(x_{n})}\nonumber\\
&=& G'_{n}(x_{1},\cdots, x_{n})
\end{eqnarray}
However, it is clear that the n-point function that appear in
equation (\ref{eq:31}) is not equal to the n-point function given
in equation (\ref{eq:24}).

Thus, we see that the method of first stochastically averaging the
classical action in the presence of the background $\tilde{f}$ and
then quantizing the theory is not the same as one that first
quantizes theory in the presence of $\tilde{f}$ (see equation
(\ref{eq:30})) and then does the stochastic averaging.

There is another subtlety in quantizing the theory even within the
framework of doing the stochastic averaging only after quantizing
the theory in the presence of $\tilde{f}$. Let us begin with
equation (\ref{eq:25}), which represents the complete generating
functional in the presence of the stochastic background
$\tilde{f}$.  This is diagrammatically represented by
\begin{eqnarray}\label{eq:32}
W[J,\tilde{f}]=\includegraphics[totalheight=0.4 cm]{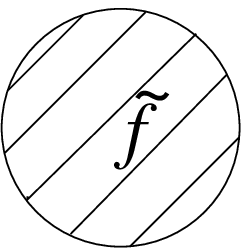}
\end{eqnarray}
while the n-point function given by equation (\ref{eq:30}) is
represented by

\begin{eqnarray}\label{eq:33}
G_{n}(x_{1}\cdots,x_{n};\tilde{f})=\includegraphics[totalheight=0.8
cm]{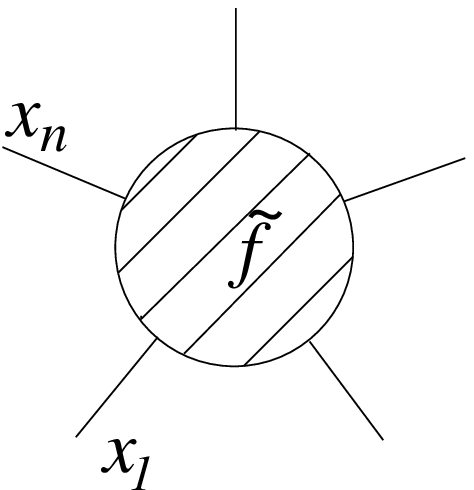}
\end{eqnarray}
Equation (31) says that
\begin{eqnarray}\label{eq:34}
\left< \includegraphics[totalheight=0.6 cm]{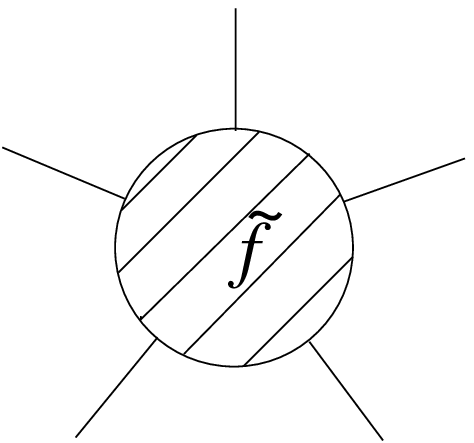}
\right>_{\tilde{f}}=\nonumber\\
\includegraphics[totalheight=0.6
cm]{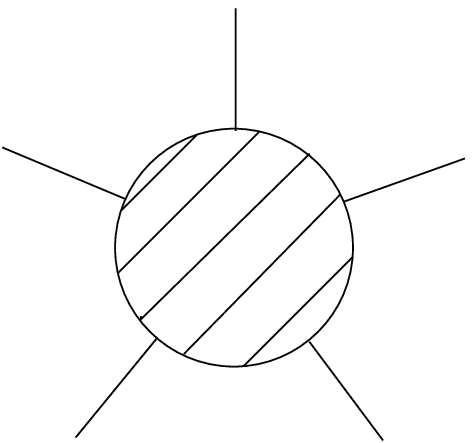}
\end{eqnarray}
where the RHS of this diagrammatic expression represents the full
n-point function from $W'_{eff}[J]$.

Following the usual field theory prescription, we define the
connected Green function generating functional via
\begin{equation}\label{eq:35}
W[J;\tilde{f}]=e^{iZ[J,\tilde{f}]}
\end{equation}
Diagrammatically, this is represented by
\begin{equation}\label{eq:36}
\includegraphics[totalheight=0.5 cm]{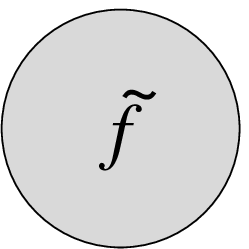}=Z[J;\tilde{f}].
\end{equation}
We compute the connected n-point Green function via
\begin{equation}\label{eq:37}
G^{c}_{n}(x_{1},\cdots,x_{n};\tilde{f})=\frac{\delta^{n}Z[J;\tilde{f}]}{\delta
J(x_{1})\cdots\delta J(x_{n})}
\end{equation}
It is clear from equation (\ref{eq:34}) that
\begin{eqnarray}\label{eq:38}
\langle Z[J;\tilde{f}]\rangle_{\tilde{f}}=\frac{1}{i}\langle \ell
n W[J;\tilde{f}]\rangle_{\tilde{f}}\nonumber\\
\neq\frac{1}{i}\ell n W'_{eff}[J]=Z'_{eff}[J]
\end{eqnarray}
where the last line of equation (\ref{eq:38}) follows from
equation (\ref{eq:26}). From equation (\ref{eq:37}), it follows
that
\begin{equation}\label{eq:39}
\langle
G^{c}_{n}(x_{1},\cdots,x_{n};\tilde{f})\rangle_{\tilde{f}}\neq
G^{'c}_{eff,n}(x_{1},\cdots,x_{n}),
\end{equation}
where the RHS is evaluated using equation (\ref{eq:26}).
Diagrammatically, this is represented by
\begin{eqnarray}\label{eq:40}
\left< \includegraphics[totalheight=0.6
cm]{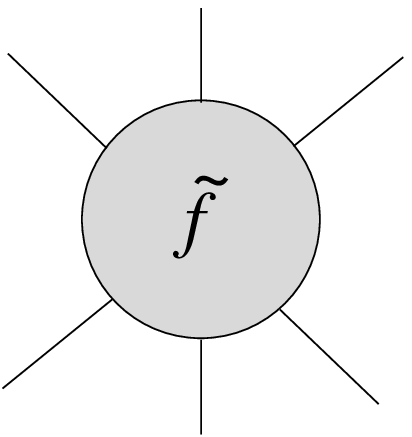}\right>_{\tilde{f}} \neq\nonumber\\
\includegraphics[totalheight=0.6 cm]{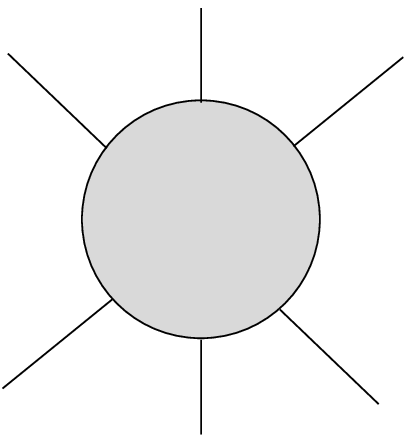}
\end{eqnarray}

If this non-equivalence is true for the connected n-point function
it is easy to show that it is also true for the
one-particle-irreducible functions derived from the
$\Gamma[\tilde{\Phi},\tilde{f}]$ and $\Gamma[\tilde{\Phi}]$, which
are derived via Legendre transformation. Define the "classical"
fields
\begin{eqnarray}
\tilde{\Phi}(\tilde{f})=\frac{\delta Z[J;\tilde{f}]}{\delta
J(x)},\label{eq:41}\\
\tilde{\Phi}_{eff}=\frac{\delta Z'_{eff}[J]}{\delta
J(x)},\label{eq:42}
\end{eqnarray}
where equation (\ref{eq:42}) makes use of equations (\ref{eq:38})
and (\ref{eq:26}) and it is obvious that
\begin{eqnarray}
\langle\tilde{\Phi}(\tilde{f})\rangle_{\tilde{f}}\neq\tilde{\Phi}_{eff}.\nonumber
\end{eqnarray}

The effective action, which is also the one-particle-irreducible
generating functional is defined by
\begin{eqnarray}
\Gamma[\tilde{\Phi},\tilde{f}]=Z[J,\tilde{f}]-\int
d^{4}x\tilde{\Phi}(\tilde{f})J,\label{eq:43}\\
\tilde{\Gamma}'_{eff}[\tilde{\Phi}_{eff}]=Z'_{eff}[J]-\int d^{4}x
J\tilde{\Phi}_{eff}.\label{eq:44}
\end{eqnarray}
where it is implied that equations (\ref{eq:41}) and (\ref{eq:42})
are inverted to solve for the sources in terms of the "classical"
fields. Again, it is clear that
\begin{equation}\label{eq:45}
\langle\tilde{\Gamma}[\tilde{\Phi},\tilde{f}]\rangle_{\tilde{f}}\neq\tilde{\Gamma}'_{eff}[\Phi_{eff}].
\end{equation}

Here, we see that although there is ambiguity when stochastic
averaging is done at the level of connected and
one-particle-irreducible generating functionals and n-point
functions, there is no such ambiguity at the level of the full
generating functional and full n-point functions. Since the
S-matrix is expressed in terms of the full n-point functions, the
quantum theory in a stochastic background is consistent and
well-defined as long as we make use of the action given by
equation (\ref{eq:28}).

\section{The effective quantum dynamics}

We begin with the path-integral in the non-linear gauge given by
\begin{eqnarray}\label{eq:46}
W&=&\int(dt^{a}_{\mu})(df^{a})(d\psi)(d\bar{\psi})\delta(\partial\cdot
t^{a}-\frac{1}{g\ell^{2}}f^{a})\delta(\rho^{a})\nonumber\\
&\cdot& det^{-4}(1+\vec{f}\cdot\vec{f})det\theta
exp.\{-(S_{YM}+S_{fermion})\}.
\end{eqnarray}
Let us rewrite the delta functionals by
\begin{eqnarray}
\delta(\partial\cdot t^{a}-\frac{1}{g\ell^{2}}f^{a})&=&
det[\frac{1}{(1+\vec{f}\cdot\vec{f})^{j}}]\delta(\frac{1}{(1+\vec{f}\cdot\vec{f})^{j}}(\partial\cdot
t^{a}-\frac{1}{g\ell^{2}}f^{a}))\label{eq:47}\\
\delta(\rho^{a})&=&
det[\frac{1}{(1+\vec{f}\cdot\vec{f})^{k}}]\delta(\frac{1}{(1+\vec{f}\cdot\vec{f})^{k}}\rho^{a})\label{eq:48}
\end{eqnarray}
where j and k are positive integers.  The reason for equations
(\ref{eq:47}) and (\ref{eq:48}) will become clear later.
Expressing the determinants in terms of ghosts, the path-integral
can be written as
\begin{equation}\label{eq:49}
W=\int(dt^{a}_{\mu})(df^{a})(d\psi)(d\bar{\psi})(du^{a})(d\bar{u}^{a})exp.\{-S'\},
\end{equation}
where
\begin{eqnarray}
S'&=& S_{YM}+S_{fermion}+S_{gf}+S_{ghosts},\label{eq:50}\\
S_{gf}&=& \int
d^{4}x\{(\frac{1}{\alpha})\frac{1}{(1+\vec{f}\cdot\vec{f})^{2j}}(\partial\cdot
t^{a}-\frac{1}{g\ell^{2}}f^{a})^{2}+\frac{1}{\beta}\frac{1}{(1+\vec{f}\cdot\vec{f})^{2k}}(\rho^{a})^{2}\},\label{eq:51}\\
S_{ghosts}&=&\int
d^{4}x\bar{u}^{a}\frac{1}{(1+\vec{f}\cdot\vec{f})^{4+j+k}}\theta^{ab}u^{b}.\label{eq:52}
\end{eqnarray}

Implementing the background decomposition given by equation
(\ref{eq:3}) in the path-integral, we get the vacuum to vacuum
functional in the background $\tilde{f}^{a}(x)$, i.e.,
\begin{equation}\label{eq:53}
W[\tilde{f}^{a}]=\int(dt^{a}_{\mu})(d\phi^{a})(d\psi)(d\bar{\psi})(du^{a}d\bar{u}^{a})exp.\{-S'\},
\end{equation}
where
$S'(t^{a}_{\mu},\psi,\bar{\psi},f^{a}=\tilde{f}^{a}+\phi^{a},\bar{u}^{a},u^{a})$.
As equations (\ref{eq:26}), (\ref{eq:27}), (\ref{eq:28}) and
(\ref{eq:34}) show, as far as the full generating functional and
the n-point Greens functions are concerned, replacing $S'$ by
$S'_{eff}$ (see equation (\ref{eq:28})) in equation (\ref{eq:53})
is equivalent to computing physical processes with
$W[\tilde{f}^{a}]$ and then doing the stochastic averages at the
end. Now we will see the reason for equations (\ref{eq:47}) and
(\ref{eq:48}). Equation (\ref{eq:28}) says we have to evaluate
\begin{eqnarray}
\langle S'\rangle_{\tilde{f}}=\langle S_{YM}\rangle +\langle
S_{fermion}\rangle + \langle S_{gf}\rangle + \langle
S_{ghosts}\rangle\label{eq:54}\\
\int
d^{4}xd^{4}y\langle(\underbrace{\mathcal{L}_{YM}+\mathcal{L}_{fermion}+\mathcal{L}_{gf}
+\mathcal{L}_{ghosts})_{x}(\mathcal{L}_{YM}+\mathcal{L}_{fermion}+\mathcal{L}_{g
f}+\mathcal{L}_{ghosts}})_{y}\rangle_{\tilde{f}},\label{eq:55}
\end{eqnarray}
etc.

The stochastic averages involve integrals of the form
\begin{equation}\label{eq:56}
\lim_{\sigma\rightarrow
0}(\pi^{-3/2}\sigma^{+3/2})\int^{\infty}_{0}\frac{r^{2m}}{(1+r^{2})^{n}}e^{-\sigma
r^{2}}dr = \left\{\begin{array}
    {r@{\quad\quad}l}
    0,& \mbox{for}\,\,m\leq n\\ & \\

    non-zero, finite,&  \mbox{for}\,m=n+1\\ & \\

    diverges,&  \mbox{for}\,m \geq 0, n+2.
    \end{array}\right.
\end{equation}

Because of the $\frac{1}{(1+\vec{f}\cdot\vec{f})^{2j}}$  term that
goes with $(\partial\cdot t^{a}-\frac{1}{g\ell^{2}}f^{a})^{2}$,
when we expand using the decomposition given by equation
(\ref{eq:3}) and by suitable choice of j, the resulting integral
will always yield zero.  The same thing is true with the term
$\sim(\rho^{2})$ and the ghost term. Thus, $\langle
S_{gf}\rangle_{\tilde{f}}=\langle
S_{ghosts}\rangle_{\tilde{f}}=0$. And $\langle
S'\rangle_{\tilde{f}}$ yields just the term given by equation
(\ref{eq:19}).

As for equation (\ref{eq:55}), all the correlated terms involving
$\mathcal{L}_{gf}$ and $\mathcal{L}_{ghosts}$ vanish for the same
reason as above. This means that $S'_{eff}$ does not involve any
ghosts $\bar{u}^{a}, u^{a}$ and we can just lump the ghosts
measure with the normalization of the path-integral.

The fermion-fermion correlated term will yield the following
non-local, four fermi interaction
\begin{equation}\label{57}
NLFF=\frac{g^{2}}{2}\int
d^{4}xd^{4}y(\bar{\psi}\gamma_{\mu}T^{a}\psi)_{x}\langle
A^{a}_{\mu}(x)A^{a'}_{\nu}(y)\rangle_{\tilde{f}}(\bar{\psi\gamma_{\nu}T^{a'}\psi)_{y}}
\end{equation}
We will evaluate the stochastic average by making use of
\begin{equation}\label{eq:58}
\partial_{\mu}A^{a}_{\mu}(x)=\frac{1}{g\ell^{2}}f^{a}(x)=\frac{1}{g\ell^{2}}\tilde{f}^{a}(x)+\frac{1}{g\ell^{2}}\phi^{a}(x).
\end{equation}

From equation (\ref{eq:58}), we must have
\begin{equation}\label{eq:59}
\partial^{x}_{\mu}\partial^{y}_{\nu}\langle
A^{a}_{\mu}(x)A^{b}_{\nu}(y)\rangle_{\tilde{f}}=\frac{1}{g^{2}\ell^{3}}\delta(x-y)+...,
\end{equation}
where $x=(x_{\mu}x_{\mu})^{1/2}$ and $y=(y_{\mu}y_{\mu})^{1/2}$.
Equation (\ref{eq:59}) implies that
\begin{equation}\label{eq:60}
\langle
A^{a}_{\mu}(x)A^{b}_{\nu}(y)\rangle=(\frac{1}{g^{2}\ell^{3}})\frac{x_{\mu}}{x}\frac{y_{\nu}}{y}\delta^{ab}|x-y|+...
\end{equation}
The equivalence follows from the fact that for a spherically
symmetric function,
$\partial^{x}_{\mu}=\frac{x_{\mu}}{x}\frac{d}{dx}$ and
$\frac{d^{2}}{dx^{2}}|x-y|=\delta(x-y)$.  Substituting equation
(\ref{eq:60}) in NLFF, we find
\begin{equation}\label{eq:61}
NLFF=\frac{1}{2}(\frac{1}{\ell^{3}})\int
d^{4}xd^{4}y(\bar{\psi}\eta_{\mu}\gamma_{\mu}
T^{a}\psi)_{x}|x-y|(\bar{\psi}\gamma_{\nu}\eta_{\nu}T^{a}\psi)_{y}+...
\end{equation}
where
\begin{eqnarray}\label{eq:62}
\eta_{\mu}=(sin\theta_{1} sin\theta_{2} sin\phi, sin\theta_{1}
sin\theta_{2} cos\phi, sin\theta_{1} cos\theta_{2},cos\theta_{1}),
\end{eqnarray}
i.e., $\eta_{\mu}$ represents the unit vectors in 4D spherical
coordinates.  If the fermion field is spherically symmetric, the
angular integration does not vanish only when
$\eta_{\mu}(\vec{x})=\pm\eta_{\mu}(\vec{y})$, i.e., the 4D vectors
are collinear.  Using

\begin{equation}
\int d
\Omega_{4}\eta_{\mu}\eta_{\nu}=\frac{\pi^{2}}{2}\delta_{\mu\nu}\label{eq:63}
\end{equation}
and $\int
d\Omega_{4}=\int^{2\pi}_{0}\int^{\pi}_{0}\int^{\pi}\sin^{2}\theta_{1}
\sin\theta_{2}d\theta_{1}d\theta_{2}d\phi=2\pi^{2}$, we find that
we can write the equation (\ref{eq:61}) as
\begin{equation}\label{eq:64}
NLFF=\frac{1}{8}(\frac{1}{\ell^{3}})\int
d^{4}xd^{4}y(\bar{\psi}\gamma_{\mu}T^{a}\psi)_{x}|\vec{x}-\vec{y}|(\bar{\psi}\gamma_{\mu}T^{a}\psi)_{y}+\cdots
\end{equation}
where the points $\vec{x}$ and $\vec{y}$ in $R^{4}$  must be
co-linear, suggesting "flux-tube" configuration.  This is
surprising because the mechanism involves spherically symmetric,
stochastic $\tilde{f}^{a}(x)$ yet the four-fermi interaction
yields a co-linear configuration with linear interaction.
Furthermore, if we neglect the extra terms in equation
(\ref{eq:64}) (which will involve quadratic terms in $\phi^{a}$),
the $\phi^{a}$ integral will yield the following delta functional
\begin{equation}\label{eq:65}
\delta(\partial_{\mu}(\bar{\psi}\gamma_{\mu}T^{a}\psi)-\frac{2g}{\ell/n}\frac{x_{\mu}}{x}(\bar{\psi}\gamma_{\mu}T^{a}\psi)).
\end{equation}
As already discussed in section II, this implies equation
(\ref{eq:20}) yielding fermion solutions given by equation
(\ref{eq:21}), which points to fermions with effective length
$\sim\Lambda^{-1}_{QCD}$ and with a linear potential as given by
equation (\ref{eq:64}).  This definitely shows quark confinement
and that when we try to pull a fermion from inside a hadron, a
colorless quark-anti quark pair (a meson), will be formed because
of the linear potential.  This is the picture of "string"
breaking, although a string was never invoked in our mechanism
(spherically symmetric, stochastic $\tilde{f}^{a}(x)$).

In summary, the effective theory for quarks and "gluons" under the
approximations we make, is given by the path-integral
\begin{eqnarray}\label{eq:66}
W&=&\int(dt^{a}_{\mu})(d\psi)(d\bar{\psi})\delta(\partial_{\mu}(\bar{\psi}
\gamma_{\mu}T^{a}\psi)-\frac{2g}{\ell/n}\frac{x_{\mu}}{x}(\bar{\psi}\gamma_{\mu}T^{a}\psi))\nonumber\\
&\times & exp.\{-S_{eff}(t^{a}_{\mu},\psi,\bar{\psi})\}
\end{eqnarray}
where
\begin{eqnarray}\label{eq:67}
S_{eff}(t^{a}_{\mu},\psi,\bar{\psi})&=&\int
d^{4}x\{\frac{1}{12}(\partial_{\mu}t^{a}_{\nu}-\partial_{\nu}t^{a}_{\mu})^{2}\nonumber\\
&+&\frac{3}{6}(\frac{n}{\ell})^{2}t^{a}_{\mu}t^{a}_{\mu}+\bar{\psi}i\gamma_{\mu}\partial_{\mu}\psi
-\frac{ig}{3}(\bar{\psi}\gamma_{\mu}T^{a}\psi)t^{a}_{\mu}\}\nonumber\\
&+&\frac{1}{8}(\frac{1}{\ell^{3}})\int
d^{4}xd^{4}y(\bar{\psi}\gamma_{\mu}T^{a}\psi)_{x}|\vec{x}-\vec{y}|(\bar{\psi}\gamma_{\mu}T^{a}\psi)_{y}.
\end{eqnarray}

\section{Conclusion}

We have derived an effective dynamics for quarks and ``gluons'' as
given in equations (\ref{eq:66}) and (\ref{eq:67}).  The effective
action clearly shows a mass gap, i.e., the ``gluons'' acquired a
mass and the quarks are confined.  The mechanism for all these is
the spherically symmetric $\tilde{f}^{a}(x)$  treated as a
stochastic variable. These vacuum configurations arise from the
non-linear regime of the non-linear gauge, which we claim is the
natural generalization in Yang-Mills theory of the Coulomb gauge
in Abelian theory.\\

\section{Acknowledgement}

This research was supported in part by the Natural Sciences
Research Institute of the University of the Philippines.



\newpage
\begin{appendix}
\section{}

Here, we will derive equation (\ref{eq:17}).  The starting point
is equation (\ref{eq:9}).  Using equation (\ref{eq:3}) in
equations (\ref{eq:10}) to (\ref{eq:16}), we get
\begin{eqnarray}
Z^{a}_{\mu\nu}(f=\tilde{f}+\phi)&=&X^{abc}(\tilde{f}+\phi)(\partial_{\mu}\tilde{f}^{b}
+\partial_{\mu}\phi^{c})(\partial_{\nu}\tilde{f}^{c}+\partial_{\nu}\phi^{c})\nonumber\\
&=&[X^{abc}(\tilde{f})+\frac{\delta X^{abc}}{\delta
f^{d}}(\tilde{f})\phi^{d}+\frac{1}{2!}\frac{\delta^{2}X^{abc}}{\delta
f^{d}\delta f^{e}}\phi^{d}\phi^{e}+\cdots]\nonumber\\
&\times&(\partial_{\mu}\tilde{f}^{b}+\partial_{\mu}\phi^{b})(\partial_{\nu}\tilde{f}^{c}+\partial_{\nu}\phi^{c})
\end{eqnarray}
Since
$X^{abc}(\tilde{f})\partial_{\mu}\tilde{f}^{b}\partial_{\nu}\tilde{f}^{c}=0$,
we find that
\begin{eqnarray}
K^{ab}_{\mu\nu}(\tilde{f})&=&2X^{cda}(\tilde{f})X^{ceb}(\tilde{f})\partial_{\alpha}\tilde{f}^{d}
\partial_{\alpha}\tilde{f}^{e}\delta_{\mu\nu}-2X^{cda}(\tilde{f})X^{ceb}(\tilde{f})\partial_{\mu}\tilde{f}^{d}\partial_{\nu}\tilde{f}^{e}\\
M^{ab}(\tilde{f})&=&\frac{\delta X^{cde}}{\delta
f^{a}}(\tilde{f})\frac{\delta X^{cfg}}{\delta
f^{b}}(\tilde{f})\partial_{\mu}\tilde{f}^{d}\partial_{\mu}\tilde{f}^{f}\partial_{\nu}\tilde{f}^{e}\partial_{\nu}\tilde{f}^{g}\\
N^{ab}_{\nu}(\tilde{f})&=& 4 X^{cda}\frac{\delta X^{cef}}{\delta
f^{b}}\partial_{\nu}\tilde{f}^{e}\partial_{\nu}\tilde{f}^{d}\partial_{\mu}\tilde{f}^{f}
\end{eqnarray}

The rest of the terms in the first bracket of equation
(\ref{eq:17}) can be arrived at by considering the second and
higher functional derivatives of $X^{abc}$

As for the second bracketted terms in equation (\ref{eq:17}),
these had been derived in reference [8].  The third bracketted
terms, which give the scalars-"gluons" interaction in a background
$\tilde{f}^{a}$, makes use of equations (\ref{eq:10}) to
(\ref{eq:16}). This is an infinite series of complicated terms.
The important point is that each term has sufficient powers of
$(1+\vec{f}\cdot\vec{f})$ in the denominator, which makes it
vanish upon stochastic averaging.

As for the fourth bracketted terms, these arise from
\begin{eqnarray}
A^{a}_{\mu}(x)&=&
R^{ab}(f)(\frac{1}{g}\partial_{\mu}f^{b}+t^{b}_{\mu})\nonumber\\
&=&R^{ab}(\tilde{f}+\phi)(\frac{1}{g}\partial_{\mu}\tilde{f}^{b}+\frac{1}{g}\partial_{\mu}\phi^{b}+t^{b}_{\mu})\nonumber\\
&=&[R^{ab}(\tilde{f})+\frac{\delta R^{ab}}{\delta f^{c}}(\tilde
{f})\phi^{c}+\frac{1}{2!}\frac{\delta^{2}R^{ab}}{\delta
f^{c}\delta f^{d}}(\tilde{f})\phi^{c}\phi^{d}+\cdots]\nonumber\\
&
&(\frac{1}{g}\partial_{\mu}\tilde{f}^{b}+\frac{1}{g}\partial_{\mu}\phi^{b}+t^{b}_{\mu}).
\end{eqnarray}

\section{}

We will derive equation (\ref{eq:19}) from equation (\ref{eq:17}).
We will make use of equation (\ref{eq:18}) and (\ref{eq:56}).  Let
us evaluate in detail one stochastic average.
\begin{eqnarray}
\langle
K^{ab}_{\mu\nu}(\tilde{f})\rangle_{\tilde{f}}&=&\mathcal{N}\int(d\tilde{f}^{a})X^{cda}(\tilde{f})X^{ceb}(\tilde{f})
(\partial_{\alpha}\tilde{f}^{d}\partial_{\alpha}\tilde{f}^{e}\delta_{\mu\nu}-\partial_{\mu}\tilde{f}^{d}\partial_{\nu}\tilde{f}^{e})\nonumber\\
&\times &
exp.\{-\frac{1}{\ell}\int^{\infty}_{0}ds\tilde{f}^{a}(s)\tilde{f}^{a}(s)\}.
\end{eqnarray}
Using equation (\ref{eq:29}) and equation (\ref{eq:13}), we find
that we need
\begin{eqnarray}
\langle \tilde{f}^{d}(x+\frac{\ell}{n})\rangle_{\tilde{f}}&=&0\\
\langle\tilde{f}^{d}(x+\frac{\ell}{n})_{\tilde{f}}(x+\frac{\ell}{n})\rangle_{\tilde{f}}&=&\frac{1}{2}\delta^{de}\lim_{\sigma\longrightarrow
0}(\frac{1}{\sigma})
\end{eqnarray}
where $\sigma=\frac{\Delta s}{\ell}$.  The stochastic average at
point x, which goes with $\frac{1}{\sigma}$ is of the form
\begin{eqnarray}
\pi^{-3/2}\sigma^{3/2}\int\frac{r^{2}drd\Omega}{(1+r^{2})^{4}}[-(1+2r^{2})\epsilon^{cda}+2\delta^{ad}x^{e}-2\delta^{ce}x^{d}\nonumber\\
+3\epsilon^{cdf}x^{f}x^{a}-3\epsilon^{caf}x^{f}x^{d}+\epsilon^{daf}x^{c}x^{f}][-(1+2r^{2})\epsilon^{ceb}\nonumber\\
+2\delta^{ce}x^{b}-2\delta^{cb}x^{e}+3\epsilon^{ceg}x^{g}x^{b}-3\epsilon^{cbg}x^{g}x^{e}+\epsilon^{ebg}x^{g}x^{c}]\nonumber\\
\times e^{-\sigma r^{2}}
\end{eqnarray}

Since the numerator only has 6 powers of r at most, while the
denominator has 8, the integral is at best
\begin{equation}
\pi^{-3/2}\sigma^{+3/2}\int^{\infty}_{0}\frac{e^{-\sigma
r^{2}}}{1+r^{2}}dr=\pi^{-3/2}\sigma^{-3/2}[1-\Phi(\sigma^{3/2})\frac{\pi}{2}e^{\sigma}]
\end{equation}
where $\Phi$ is the error function (not to be confused with the
field that collectively represents $t^{a}_{\mu},\phi^{a}, \phi,
\bar{\psi}$ in section III) with expansion
\begin{equation}
\Phi(\sigma^{1/2})=\frac{2}{\sqrt{\pi}}[\sigma^{1/2}-\frac{1}{3}\sigma^{3/2}+\frac{1}{10}\sigma^{5/2}+\cdots]
\end{equation}
Thus, this integral behaves, at best, like $\sigma^{+3/2}$, as
$\sigma\longrightarrow 0$. Combining this with the
$\frac{1}{\sigma}$ factor, we find this term to vanish like
$\sigma^{1/2}$. There is another term in $\langle
K^{ab}_{\mu\nu}\rangle_{\tilde{f}}$ with
$\tilde{f}^{d}(x)\tilde{f}^{e}(x)$ of the derivatives, which will
combine with the rest given in (B.4). But this term will yield a
stochastic average which will vanish like $\sigma$ as
$\sigma\longrightarrow 0$. Thus, we find
\begin{equation}
\langle K^{ab}_{\mu\nu}(\tilde{f})\rangle_{\tilde{f}}=0.
\end{equation}
Using similar analysis, it is trivial to show that all the other
stochastic averages in the first and 3rd brackets of equation
(\ref{eq:17}) vanish.  The only non-vanishing stochastic averages
are
\begin{eqnarray}
\langle
\mathbb{R}^{ab}(\tilde{f})\rangle_{\tilde{f}}=\frac{1}{3}\delta^{ab}\\
\langle\mathbb{Y}^{ab}\rangle_{\tilde{f}}=\delta^{ab}(\frac{3}{4})(\frac{n^{2}}{\ell^{2}})
\end{eqnarray}
These averages yield the first and second terms of equation
(\ref{eq:19}).

Finally, from equation (A.5), we get
\begin{equation}
\langle
A^{a}_{\mu}\rangle_{\tilde{f}}=\frac{1}{3}t^{a}_{\mu}+\frac{1}{3g}\partial_{\mu}\phi^{a}
-\frac{2}{3g}(\frac{n}{\ell})\frac{x_{\mu}}{x}\phi^{a}.
\end{equation}
\end{appendix}
\end{document}